\documentclass{aa}
\usepackage[varg]{txfonts}

\usepackage{graphicx}
\usepackage{amsmath,amssymb}
\usepackage{xcolor}
\usepackage{enumitem}
\usepackage{bbm}
\usepackage{xspace}
\usepackage[export]{adjustbox}

\usepackage{natbib}
\bibpunct{(}{)}{;}{a}{}{,}
\providecommand{\sorthelp}[1]{}

\newcommand{\lollipop}{\textsc{LoLLiPoP}}
\newcommand{\hillipop}{\textsc{HiLLiPoP}}
\newcommand{\planck}{\textit{Planck}\xspace}

\usepackage[colorlinks=true, allcolors=blue]{hyperref}

\begin{document}

\title{Reconstructing the epoch of reionisation with Planck PR4}

\author{
   S.~Ili\'c \inst{1}
   \and
   M.~Tristram \inst{1}
   \and
   M.~Douspis \inst{2}
   \and
   A.~Gorce \inst{2}
   \and
   S.~Henrot-Versill\'e \inst{1}
   \and
   L.~T.~Hergt \inst{1}
   \and
   M.~Langer \inst{2}
   \and
   L.~McBride \inst{2}
   \and
   M.~Muñoz-Echeverr\'{i}a \inst{3}
   \and
   E.~Pointecouteau \inst{3}
   \and
   L.~Salvati \inst{2}
}

\institute{
IJCLab, Universit\'e Paris-Saclay, CNRS/IN2P3, IJCLab, 91405 Orsay, France
\label{1}
\and
Universit\'{e} Paris-Saclay, CNRS, Institut d'Astrophysique Spatiale, 91405, Orsay, France
\label{2}
\and
IRAP, CNRS, Universit\'e de Toulouse, CNES, UT3-UPS, (Toulouse), France
\label{3}
}

\date{\today}

\abstract{
    The epoch of reionisation is a key phase in cosmic history, characterised by the ionisation of the intergalactic medium by the first luminous sources. In this work, we constrain the reionisation history of the Universe using data from the cosmic microwave background, more specifically the latest \planck Public Release 4 (PR4) dataset. We investigate a wide range of reionisation models, from simple parametric descriptions to more flexible non-parametric approaches, systematically evaluating their impact on the inferred constraints. Special attention is given to implicit priors introduced by each model and their influence on the derived reionisation optical depth, $\tau$. To achieve this, we employ both Bayesian and frequentist methods to derive robust constraints. We obtain consistent estimates of $\tau$ across models, highlighting the robustness of the constraints on the integrated optical depth derived from the \planck PR4 data. Averaging across models, the posterior means and best-fit values, respectively, yield $\tau = 0.0576 \pm 0.0060$ and $\tau = 0.0581$, highlighting the presence of small volume effects. Based on our analysis, we estimate that an additional uncertainty, associated with the modelling of reionisation, contributes an error of approximately $\sigma_\tau\!\sim\!0.0006$. Beyond the integrated optical depth, our analysis reveals that the inferred ionisation fraction as a function of redshift is highly model-dependent. While current CMB data do not favour significant early ionisation, they are consistent with a modest contribution from ionised gas at very early times ($z>15$). Although indicative upper bounds can be placed on such contributions, these limits remain strongly dependent on the assumed model.
}


\maketitle

\section{Introduction}
\label{sec:Introduction}

The epoch of reionisation (EoR) represents a critical phase in cosmic history during which the intergalactic medium transitioned from a neutral to ionised state due to the emergence of the earliest sources of ionising light. This process is estimated to have occurred between redshifts $z \sim 20$ and $z \sim 6$, corresponding to approximately 150 million to 1 billion years after the Big Bang \citep{2001PhR...349..125B,2013fgu..book.....L}, during which the Universe transitioned from the so-called Cosmic Dark Ages to the highly structured and luminous cosmos we observe today. Understanding the EoR is crucial for cosmic structure formation theories and offers insight into the properties of the first star and galaxy populations in the early Universe.

Observational constraints from the cosmic microwave background (CMB), particularly measurements of the optical depth, $\tau$, from the \planck satellite, now yielding $\tau<0.06$, suggest that reionisation was a relatively recent process likely driven by the first galaxies and possibly by population III stars \citep{planck2014-a25}. Additional insights come from high-redshift quasar spectra, which exhibit Gunn-Peterson troughs indicative of a fully ionised intergalactic medium by $z \sim 5.4$ \citep{2013ApJ...779...24V,2015PASA...32...45B,BosmanDavies_2022}. The evolution of Lyman-alpha emitters also provides evidence of a patchy and inhomogeneous reionisation process \citep{2018ApJ...856....2M}. Recent observations from the James Webb Space Telescope \citep[JWST;][]{Gardner:2006} have dramatically improved our understanding of galaxy populations during the EoR, revealing an unexpectedly high abundance of ultraviolet-luminous galaxies at redshifts $z>9$ \citep{Harikane:2023,Bouwens:2023}. These observations extend the ultraviolet luminosity function (UVLF) out to $z\sim13$, particularly at the bright end. Moreover, JWST has uncovered surprisingly massive and evolved galaxies at $z>10$, challenging existing models of early galaxy formation and suggesting a more rapid and efficient assembly of stellar mass than previously anticipated \citep{Labbe:2023}. If these galaxies formed early and were efficient producers of ionising radiation, the reionisation of the Universe may have progressed more rapidly than inferred from CMB data, potentially implying a higher optical depth, up to $\tau \gtrsim 0.07$ \citep{Munoz:2024}. Reconciling these findings with CMB-based constraints is essential for developing a coherent picture of the timeline and sources of cosmic reionisation.

In this context, the present work aims at providing a comprehensive study of the constraints on the EoR using CMB data, which encode imprints of the ionisation history through its polarisation and temperature anisotropies. We explore a broad range of reionisation models -- from simple parametric forms to more flexible non-parametric descriptions -- to evaluate the impact of model assumptions on the inferred constraints. Our analysis focuses primarily on the reionisation optical depth, $\tau$, but also extends to the reconstruction of the ionisation history as a function of redshift, $x_{\rm e}(z)$. We pay particular attention to the implicit priors introduced by each model on the reionisation history and key parameters such as $\tau$. 

The \planck satellite has provided some of the most detailed CMB measurements to date on the full sky, enabling stringent constraints on the parameters of our cosmological model, including the reionisation history \citep{planck2014-a25}.  In this work, we use the latest data from the \planck mission -- Public Release 4 (PR4) -- to constrain various reionisation models. The PR4 dataset represents an improvement in data processing and error analysis \citep{planck2020-LVII}, offering more precise measurements on large scales, especially in polarisation and tighter constraints on cosmological parameters \citep{Rosenberg:2022sdy, 2024A&A...682A..37T}.

This manuscript proceeds by presenting first in Sect.~\ref{sec:Models} the range of considered reionisation models to be confronted to our selected datasets. The latter are described in Sect.~\ref{sec:Data}, alongside our methodology for deriving cosmological constraints. Our results on the reionisation optical depth are detailed in Sect.~\ref{sec:results_tau} and the interpretation in terms of the reionisation history (and other derived quantities) is discussed in Sect.~\ref{sec:results_xe}. We summarise our findings and perspectives in Sect.~\ref{sec:Conclusions}.

\section{Modelling and probing the reionisation history}
\label{sec:Models}

Previous measurements of the CMB anisotropies by the \planck satellite \citep{planck2014-a25} have provided substantial advancements in our understanding of the EoR, improving our estimates of its timing and duration. Most notably, these observations have been key to constraining the optical depth due to Thomson scattering, usually denoted $\tau$, a critical parameter in the $\Lambda$CDM cosmological model:
\begin{equation}
    \tau = \int_{0}^{z_{\rm max}} \sigma_{\rm T} n_{\rm e}(z) \frac{dr}{dz} dz \, ,
    \label{eq:tau_def}
\end{equation}
where $\sigma_{\rm T}$ is the Thomson cross-section, $n_{\rm e}(z)$ is the volume-averaged free electron (proper) number density, $z_{\rm max}$ is an arbitrary redshift upper bound (high enough to encompass the whole EoR), and $dr/dz$ is the line-of-sight (proper) distance per unit redshift (throughout we assume that the Universe is flat, in which case $dr/dz = c/(1 + z)/H(z)$, with $H(z)$ the expansion rate).

Thomson scattering between the CMB photons and free electrons generates linear polarisation from the quadrupole moment of the CMB radiation field at the scattering epoch. Such scattering occurs whenever a significant amount of free electrons is present in the Universe, namely at recombination as well as during the EoR. For the latter, the probability of scattering is directly linked to $\tau$ (namely $1-e^{-\tau}$). Such rescattering of the CMB photons during reionisation then generates additional polarisation anisotropies at large angular scales.

This signal, appearing primarily as a low-$\ell$ bump in the E-mode polarisation angular power spectrum, is linked to the horizon size at the ``new last-rescattering surface'' and thus depends on the redshift of reionisation. The amplitude of the bump scales roughly as the square of the reionisation optical depth, $\tau$, but its exact shape encodes information about the evolution of the ionisation fraction as a function of redshift~\citep{Zaldarriaga:1997,Kaplinghat:2003}. At the same time, the rescattering of CMB photons also reduces the overall amplitude of all scalar power spectra, scaling as $e^{-2\tau}$. This introduces a correlation between $\tau$ and the primordial amplitude of the scalar perturbations, $A_{\rm s}$, when inferring their values from observational data, as the latter is sensitive to the overall amplitude of the power spectra which scales as $A_{\rm s} e^{-2\tau}$. Additionally, the kinetic Sunyaev-Zel'dovich (kSZ) effect, which arises from the scattering of CMB photons off moving electrons in ionised regions, provides complementary constraints on the duration and patchiness of reionisation through small-scale CMB anisotropies \citep{McQuinn:2005,Mesinger:2012,GorceIlic_2020}. However, as its modelling requires additional assumptions and lies beyond the scope of this work, we do not include kSZ-based constraints in our analysis.

\begin{figure}[!t]
    \center
    \includegraphics[width=.95\columnwidth]{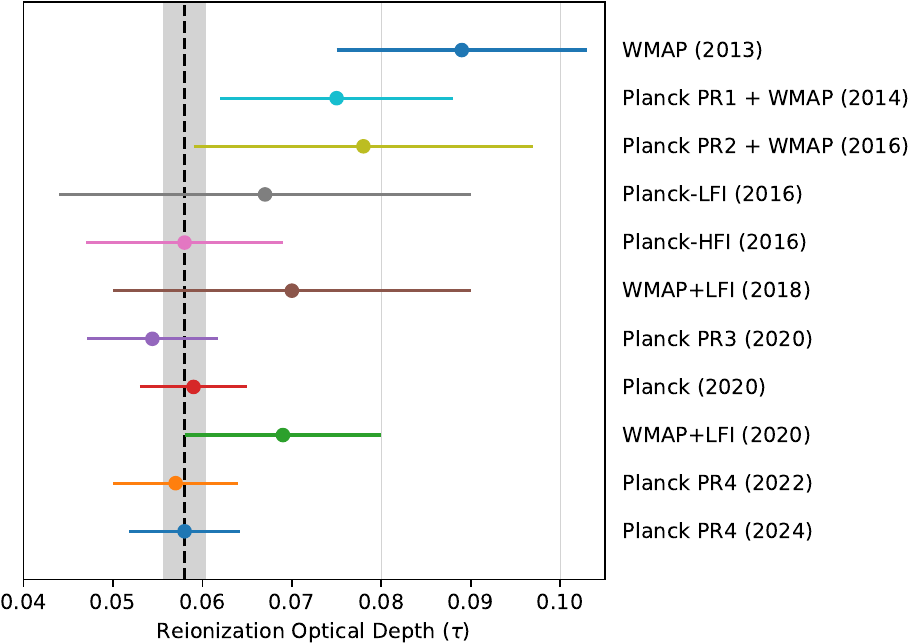}
    \caption{Mean values (filled circles) and 68\% credible intervals (horizontal bars) obtained from posterior distributions of $\tau$, derived from CMB power spectra measurements including:
    WMAP 2013 \citep{2013ApJS..208...19H},
    Planck PR1 + WMAP 2014 \citep{planck2013-p08},
    Planck PR2 + WMAP 2016 \citep{planck2014-a15},
    Planck-LFI 2016 \citep{planck2014-a15},
    Planck-HFI 2016 \citep{planck2014-a25},
    WMAP+LFI 2018 \citep{2018ApJ...863..161W},
    Planck PR3 2020 \citep{planck2016-l06},
    Planck 2020 \citep{2020A&A...635A..99P},
    WMAP+LFI 2022 \citep{2023A&A...675A..12P},
    Planck PR4 2022 \citep{planck2020-LVII}.
    The shaded grey region highlights the cosmic variance limit on the measurement of $\tau$, centred on the value from \planck PR4 2024 \citep{2024A&A...682A..37T}.
    }
    \label{fig:tau-history}
\end{figure}

Accurate measurements of $\tau$ through the CMB are challenging due to foreground contamination and instrumental systematic uncertainties. These effects are particularly significant on large angular scales, from where most of the constraints on $\tau$ originate. The first estimates by WMAP suggested a high $\tau$ \citep{2003ApJS..148..161K} corresponding generally to an early start for reionisation processes, but subsequent measurements by \planck adjusted $\tau$ downwards (see Fig.~\ref{fig:tau-history}), thanks to a re-assessment of the dust contamination in particular~\citep{planck2014-a25,planck2016-l06}. This shift restored consistency with other astrophysical data, suggesting a later reionisation not requiring exotic early-time ionising sources. Further refinements in the analysis of the \planck PR4 yielded an even better sensitivity in polarisation measurements, achieving a precision now reaching $\sim$2.5 times the cosmic-variance limited uncertainty~\citep{2024A&A...682A..37T}.

However, it is important to highlight that the aforementioned CMB measurements were derived under the assumption of a simple reionisation history, modelled as a single-step transition in the ionisation fraction, $x_{\rm e}(z)$, using a hyperbolic tangent function. While alternative approaches -- such as principal component analysis \citep{Heinrich:2021}, FlexKnot \citep{planck2016-l06}, and Poly-reion \citep{Paoletti:2020} -- have been explored using \planck 2018 data, it should be emphasised that the low-$\ell$ $EE$ likelihood presented in \citet{planck2016-l05} is an empirical likelihood based on simulations constructed within a fixed $\Lambda$CDM framework assuming a hyperbolic tangent reionisation. As such, it is not suitable for constraining more extended or flexible reionisation scenarios.

In the present work, we set out to test various reionisation scenarios by constraining the evolution of the fraction of ionised matter as a function of redshift, thereafter $x_{\rm e}(z)$, defined as the ratio of $n_{\rm e}(z)$ over the hydrogen number density $n_{\rm H}(z)$ and ranging between $0 \leq x_{\rm e}(z) \leq (1+2f_{\rm He})$ (where $f_{\rm He}=n_{\rm He}/n_{\rm H}$ is the primordial helium-to-hydrogen nucleon ratio). This process involves confronting observational data to models with various levels of sophistication, ranging from instantaneous to extended, from symmetric to asymmetric, and from physically motivated to model-independent. The variety of reionisation models comes from the recognition that different, still mostly unknown mechanisms -- ranging from the first stars and galaxies to quasars -- may have their own distinct signatures, at distinct moments on the reionisation timeline.

\begin{figure*}[!ht]
    \centering
    \includegraphics[width=0.9\textwidth]{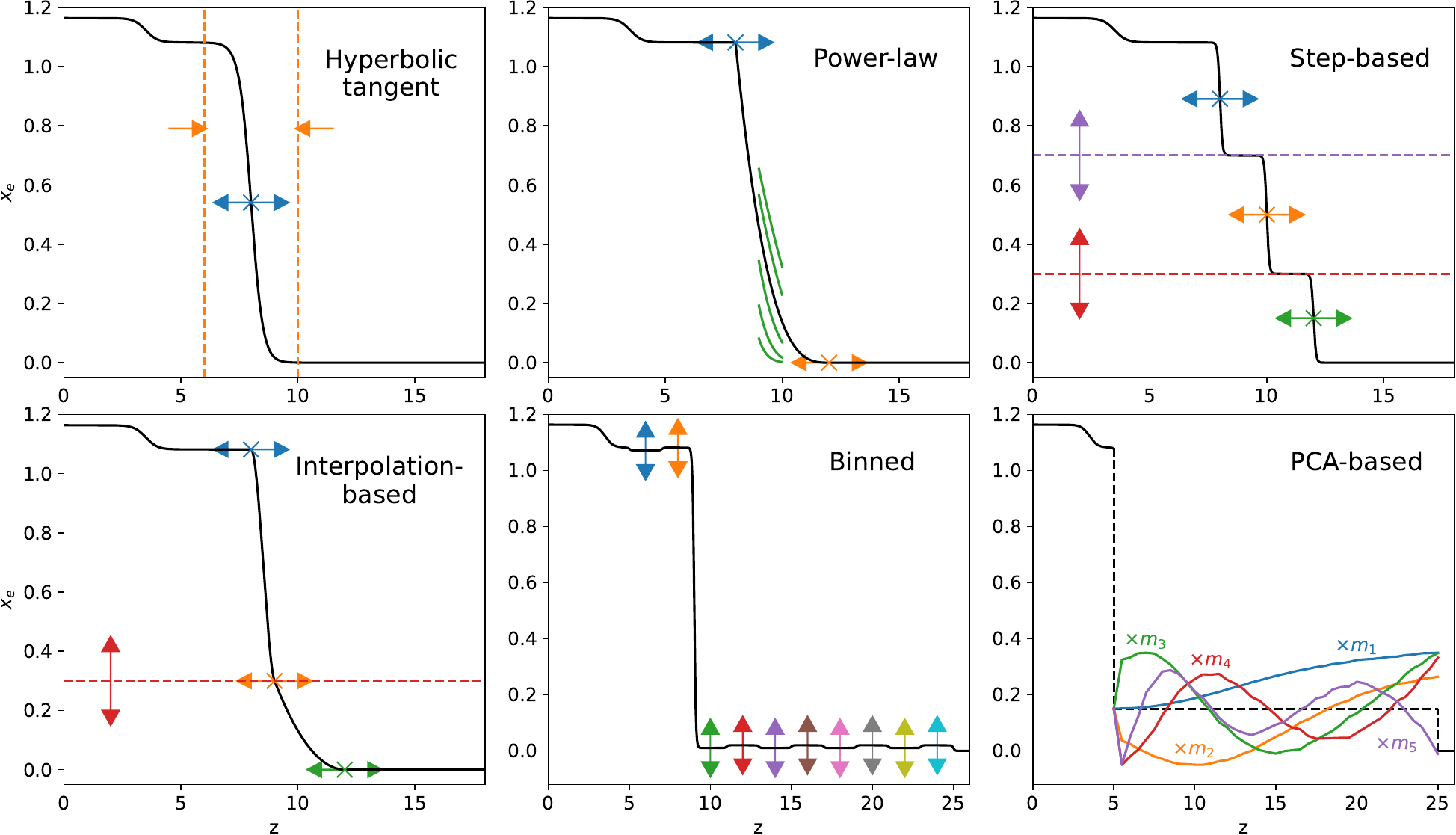}
    \caption{Graphical summary of the reionisation models considered in the present study, illustrating with coloured arrows and lines their respective degrees of freedom in modelling the history of the ionising fraction. The models are grouped by their general approach: hyperbolic tangents, power-laws, step functions, interpolations, bins, and PCA. The models are described in detail in Sect.~\ref{sec:Models}.}
    \label{fig:models-summary}
\end{figure*}

In the following, we briefly describe the suite of parametric and non-parametric models that we examine, and provide a visual summary in Fig.~\ref{fig:models-summary}. A list of all model parameters and the corresponding prior choices are summarised in Table~\ref{tab:pars_and_priors}. We note that, unless specified otherwise, we include in all models a second reionisation of helium between redshift 3 and 4 (constructed using a simple $\tanh$ function of width $\Delta z_{\rm re}^{\rm He} = 0.5$ and central redshift $z_{\rm re}^{\rm He} = 3.5$)\footnote{The precise onset and duration of the second helium reionisation remain uncertain due to limited observational data and model dependencies~\citep[e.g.][]{2016ApJ...825..144W}. However, its contribution to the optical depth is only $\delta\tau\sim0.001$, making it subdominant compared to the hydrogen and first helium reionisation.}, which is not mentioned in the model descriptions. Finally, in all our considered models, we set the beginning of reionisation at redshift 25 at the earliest (sufficiently deep to allow for an early start) and ensure that the hydrogen and first helium reionisation are ended by redshift 5 in accordance with astrophysical constraints (see references in Sect.~\ref{sec:Introduction}).

\subsection{Hyperbolic tangent models}

The classic hyperbolic tangent model features a rapid transition of the Universe from a neutral to ionised state, and has been extensively utilised in CMB analyses as the default model implemented most notably in widespread Boltzmann codes \citep[e.g. CAMB,][]{Lewis:2008}. It is parametrised by a midpoint redshift, $z_{\rm re}$, and a duration, $\Delta z_{\rm re}$, for the transition, as 
\begin{equation}
    x_{\rm e}(z) \propto \frac{1}{2} \left[ 1 + \tanh\left(\frac{1}{\Delta z_{\rm re}} \frac{(1+z_{\rm re})^\gamma - (1+z)^\gamma}{\gamma(1+z_{\rm re})^{\gamma-1}} \right) \right] \, ,
    \label{eq:tanh}
\end{equation}
where the exponent, $\gamma$, is typically set to $1.5$. We consider two forms of this model: one with a fixed duration of $\Delta z_{\rm re}=0.5$ and the other where the duration is allowed to vary (see the upper-left panel in Fig.~\ref{fig:models-summary}), enhancing the flexibility of the model to account for more or less extended reionisations, although those rarely reflect actual expected physical evolutions. We denote these two models \texttt{tanh} and \texttt{tanhdz}, respectively. We note that \texttt{tanhdz} is unique in our suite of models, as it is the only one that allows for unfinished reionisations at $z=0$.

\subsection{Power-law models}

A significant concern regarding the use of the hyperbolic tangent as a model relates to its lack of agreement with state-of-the-art hydrodynamical simulations of the EoR \citep{2015MNRAS.454.1012A,2019MNRAS.487.5739S,2023MNRAS.519.6162P,2024A&A...683A..24M}. Indeed, these simulations tend to reveal a more intricate picture, with for instance a reionisation that starts slowly with the formation of the first stars and culminates abruptly. A more physically motivated parameterisation of the EoR was first proposed and explored in \citet{Douspis:2015nca} and consequently used in the EoR analysis of the \planck data \citep{planck2014-a25}, albeit in a simplified form, which we adopt here:
\begin{equation}
    x_{\rm e}(z) =
    \begin{cases}
      (1+f_{\rm He}) & \mbox{for } z \leq z_{\rm end} \, , \\
      (1+f_{\rm He}) \left(\frac{z_{\rm beg}-z}{z_{\rm beg}-z_{\rm end}}\right)^\alpha
        & \mbox{for }z>z_{\rm end} \, .
    \end{cases}
    \label{eq:asymm}
\end{equation}

This model, by design, incorporates an element of asymmetry absent from the hyperbolic tangent via the exponent, $\alpha$, and the choice of the redshift of the start~($z_{\rm beg}$) and end~($z_{\rm end}$) of reionisation. It can accommodate a diverse range of astrophysically-plausible scenarios, which may include either a protracted onset prior to a swift completion or the inverse sequence. Two variants of the model are also considered, where the start redshift is either fixed to $z_{\rm beg}=20$ \citep[as in][]{planck2014-a25} or free (see the upper-central panel in Fig.~\ref{fig:models-summary}). We denote these models \texttt{2-pow} and \texttt{3-pow}, respectively.

\subsection{Step-based models}

We also explore in parallel a set of models based on step-like, sharp ($\Delta z=0.1$) hyperbolic tangent transitions (see the upper-right panel in Fig.~\ref{fig:models-summary}). More specifically, we adopt two models constructed out of 2 and 3 transitions, where the position of each transition is allowed to vary, as well as the height of the plateaus between transitions. This results in two models with 3 and 5 degrees of freedom, respectively, which we dub \texttt{2-steps} \citep[as used in][]{Watts:2020} and \texttt{3-steps} in the following.

\subsection{Interpolation-based models}
Another approach we consider here involves flexible histories, namely via interpolation-based models in which the ionisation fraction is interpolated across a series of user-specified $\{z_i, x_{{\rm e},i}\}$ points, called ``knots'' in the following. Such a model is simple yet versatile, allowing for the construction of a wide variety of reionisation trajectories by adjusting the specified points. We consider two types of interpolation: a simple linear interpolation model, and the ``flexknot'' model of \citet{Millea:2018}. The latter employs Piecewise Cubic Hermite Interpolating Polynomials (PCHIP), which ensure monotonic transitions between nodes and provide a smooth reionisation history (see lower-left panel in Fig.~\ref{fig:models-summary}). We consider models with 2 and 3 knots, and we note that the ionised fraction of the lowest and highest redshift nodes are always fixed, respectively, to $1+f_{\rm He}$ (to ensure reionisation is finished) and to its residual value from recombination (for continuity). We denote these models \texttt{2-knots} and \texttt{3-knots} for the linear interpolation, and \texttt{2-flex} and \texttt{3-flex} for the PCHIP interpolation, respectively.

\subsection{Binned models}
\label{sec:Models:bin}

One of the most general model-independent approaches to study the reionisation history is to only consider the average ionisation fraction over a succession of redshift intervals. In practice, this can be achieved by adopting a binned history of the ionisation fraction, which we implement through a series of sharp ($\Delta z=0.1$) consecutive hyperbolic tangent transitions with constant plateaus in-between, each corresponding to distinct, fixed redshift intervals. This discretised approach enables notably the investigation of non-monotonic reionisation histories, allowing for periods of both increased and decreased ionisation fraction within the EoR, but also more generally allows for a model-independent exploration, mostly independent of any assumption or model prior.

In the present study, we settle on a 10-bin model, spanning $z=5$ to $z=25$ and with a bin of width 2 in redshift -- our model with the seemingly highest flexibility in our analysis, which we dub \texttt{10-bins} in the following.

\subsection{Principal component analysis model}
\label{sec:Models:pca}

The principal component analysis (PCA) model offers a generic approach to reconstruct the reionisation history from CMB data, without assuming any functional form but instead using a set of complex, data-driven functions. It involves decomposing the ionisation fraction history, $x_{\rm e}(z)$, into a set of orthonormal eigenfunctions or principal components \citep{2003PhRvD..68b3001H}, which are specifically designed to maximise the sensitivity of the model to the underlying data.

In the PCA model, the ionisation fraction is then written as
\begin{equation}
    x_{\rm e}(z) = x_{\rm e}^{\rm fid}(z) + \sum_{i} m_i S_i(z) \, ,
\end{equation}
where $x_{\rm e}^{\rm fid}(z)$ denotes an arbitrary chosen fiducial ionisation history, $m_i$ represents the amplitudes of the principal components, and $S_i(z)$ are the principal component templates. These templates are constructed to satisfy several properties, namely: forming a special orthonormal basis over the redshift range of interest, minimising the covariance of the amplitude parameters given cosmic-variance-limited CMB data, and being ordered by increasing uncertainty, ensuring that the most constrained modes are considered first. This construction allows the PCA model to flexibly adapt to various possible reionisation histories without being overly prescriptive. This model has been applied to WMAP \citep{Mortonson:2008} in the past, as well as to \planck data \citep{Heinrich:2017,Heinrich:2021}.

In the present work, we built a PCA model with a set of basis functions spanning the redshift range $z=5$ to $z=25$, and a fiducial model $x_{\rm e}^{\rm fid}=0.15$ (corresponding to $\tau_{\rm fid}=0.066$, reasonably close to current estimates). The components are derived from a Fisher matrix-based approach, assuming a \planck best-fit cosmology for an all-sky, cosmic variance-limited experiment. By construction, due to the finite number of components used, the ionisation fraction, $x_{\rm e}(z)$, reconstructed from the PCA is not inherently restricted to physical bounds (i.e. between 0 and 1+$f_{\rm He}$). We keep only the first 5 components which according to \citet{Mortonson:2008} are sufficient to describe all the information on $x_{\rm e}$ carried by $C_\ell^{EE}$ up to the cosmic variance limit. During Markov chains Monte Carlo (MCMC) exploration, we impose priors to the amplitude of each mode following the prescription of \citet{Mortonson:2008}, ensuring that reionisation histories with unphysical optical depths over a broad redshift range are treated as unphysical models\footnote{We note that \citet{Millea:2018} have pointed out that this choice of prior is not optimal for the reconstruction of $\tau$. However, we can correct for its informative effect on $\tau$ using the procedure described in Sect.\ref{sec:priors}.}. We denote this model \texttt{pca-5} in the following.

\begin{table}[!ht]
    \renewcommand{\arraystretch}{1.1}
    \caption{Parameters of the considered reionisation models and their associated priors or fixed values}
    \centering
    \begin{tabular}{ccc}
    \hline
    \multicolumn{1}{c}{Parameters} & \multicolumn{2}{c}{Priors or values}          \\ \hline
    \multicolumn{3}{c}{Hyperbolic tangent models}                                  \\ \hline
                         & {\texttt{tanh}} & {\texttt{tanhdz}}                     \\
    $z_{\rm re}$         & \multicolumn{2}{c}{$\mathcal{U}(5, 25)$}                \\
    $\Delta z_{\rm re}$  & {=0.5}     & {$\mathcal{U}(0.1, 5)$}                    \\ \hline
    \multicolumn{3}{c}{Power-law models}                                           \\ \hline
                         & {\texttt{2-pow}} & {\texttt{3-pow}}                     \\
    $z_{\rm end}$        & {$\mathcal{U}(5, 20)$}     & {$\mathcal{U}(5, 25)$}     \\
    $\alpha$             & \multicolumn{2}{c}{$\mathcal{U}(0, 20)$}                \\
    $z_{\rm beg}$        & {=20}     & {$\mathcal{U}(5, 25)$}                      \\ \hline
    \multicolumn{3}{c}{Step-based models}                                          \\ \hline
                         & {\texttt{2-steps}} & {\texttt{3-steps}}                 \\
    $z_1$                & \multicolumn{2}{c}{$\mathcal{U}(5, 25)$}                \\
    $z_2$                & \multicolumn{2}{c}{$\mathcal{U}(5, 25)$}                \\
    $z_3$                & {--}     & {$\mathcal{U}(5, 25)$}                       \\
    $x_{{\rm e},2}$            & \multicolumn{2}{c}{$\mathcal{U}(0, 1+f_{\rm He})$}      \\
    $x_{{\rm e},3}$            & {--}     & {$\mathcal{U}(0, 1+f_{\rm He})$}             \\ \hline
    \multicolumn{3}{c}{Interpolation-based models}                                 \\ \hline
                         & {\texttt{2-knots}} & {\texttt{3-knots}}                 \\
                         & {\texttt{2-flex}} & {\texttt{3-flex}}                   \\
    $z_1$                & \multicolumn{2}{c}{$\mathcal{U}(5, 25)$}                \\
    $z_2$                & \multicolumn{2}{c}{$\mathcal{U}(5, 25)$}                \\
    $z_3$                & {--}     & {$\mathcal{U}(5, 25)$}                       \\
    $x_{{\rm e},2}$            & {--}     & {$\mathcal{U}(0, 1+f_{\rm He})$}             \\ \hline
    \multicolumn{3}{c}{Binned models}                                    \\ \hline
                         & \multicolumn{2}{c}{\texttt{10-bins}} \\
    $x_{{\rm e},i\in[1,10]}$     & \multicolumn{2}{c}{{$\mathcal{U}(0, 1+f_{\rm He})$}}  \\ \hline
    \multicolumn{3}{c}{PCA model}                                 \\ \hline
    $m_{i\in[1,5]}$                & \multicolumn{2}{c}{Very large uniform priors}                \\
                  & \multicolumn{2}{c}{$+$ physicality prior on $x_{\rm e}(z)$}                \\ \hline
    \end{tabular}
    \tablefoot{The notation $\mathcal{U}(a, b)$ denotes a uniform prior between $a$ and $b$.}
    \label{tab:pars_and_priors}
\end{table}

\section{Data and methodology}
\label{sec:Data}

The following section outlines the specific datasets and the computational tools used in the present analysis to test the reionisation histories described in the previous section.

\subsection{Data}

The primary dataset used in this analysis is derived from the latest and final \planck PR4 maps. These maps have been processed through the NPIPE pipeline \citep{planck2020-LVII} which includes data from the \planck Low-Frequency Instrument (LFI) and the High-Frequency Instrument (HFI) in both temperature (T) and polarisation (P). The NPIPE processing pipeline improves upon previous releases by incorporating data from the repointing periods and by providing lower noise levels and reduced systematics across all angular scales, enhancing the internal consistency between frequencies. Additionally, the PR4 includes a comprehensive set of ``End-to-End'' Monte Carlo simulations aiming to help characterise potential biases and uncertainties associated with the data processing.

To ensure unbiased estimates of the angular power spectra, cross-correlations of two independent splits of the data -- referred to as detsets -- were performed, both across frequencies and across T and P signals. Detset maps comprise subsets of detectors that are combined to produce maps with nearly independent noise properties, crucial for robust cosmological parameter estimation.

In order to extract cosmological information from the \planck PR4 data, we split the dataset in two likelihoods: one for the large angular scales (low $\ell$) and the other for small angular scales (high $\ell$). The details of these respective likelihoods are briefly described in the following.

\subsubsection{\lollipop}

\lollipop\ (Low-$\ell$ Likelihood on Polarised Power-spectra) is a likelihood focused on the low multipoles of the CMB polarisation, including the $EE$, $BB$, and $EB$ spectra. Initially applied to \planck PR3 data for reionisation studies, \lollipop\ has been upgraded to PR4 as described in \citet{2021A&A...647A.128T,2022PhRvD.105h3524T}. \lollipop\ uses cross-power spectra calculated from component-separated CMB detset maps processed by the Commander code \citep{2008ApJ...676...10E} using all available polarised frequencies from 30 to 353\,GHz.

This likelihood utilises the Hammimeche--Lewis approximation, modified to apply to cross-power spectra \citep{2008PhRvD..77j3013H,Mangilli:2015}. This method ensures a nearly Gaussian distribution of a transformed variable, derived from the measured and model power spectra. The likelihood function incorporates the covariance matrix derived from Monte Carlo simulations, ensuring that uncertainties from the CMB sample variance, statistical noise, and systematics are accurately propagated. Finally, the likelihood is marginalised over the unknown true covariance matrix \citep[as proposed in][]{2016MNRAS.456L.132S} in order to propagate the uncertainty in the estimation of the covariance matrix caused by a limited number of simulations.

In this work, we use \lollipop\ to cover the multipole range from $\ell = 2$ to $\ell = 30$ for the $EE$ spectrum, providing crucial constraints on the reionisation optical depth, $\tau$.

\subsubsection{\hillipop}

\hillipop\ (High-$\ell$ Likelihood on Polarised Power-spectra) is designed for analysing the small-scale angular power spectra ($\ell > 30$) from the \planck PR4 data. \hillipop\ uses a multi-frequency Gaussian likelihood approximation, incorporating sky maps at 100, 143, and 217 GHz, with the 353 GHz channel used to model dust contamination. This likelihood has been updated to V4.2 for the PR4 data release, utilising a larger sky fraction (up to 75\%) and refined models for foreground components such as dust and point sources \citep{2024A&A...682A..37T}.

Small-scale power spectra in $TT$, $TE$, and $EE$ are computed using Xpol \citep[an extension to polarisation of the Xspect pseudo-$C_\ell$ method, described in][]{2005MNRAS.358..833T}, which corrects for beam effects and normalises the power spectra with a mode-mixing matrix. \hillipop\ incorporates a comprehensive semi-analytic covariance matrix that includes contributions from cosmic variance, instrumental noise, and the residual foregrounds.

\subsubsection{Combining likelihoods}

By combining \lollipop\ and \hillipop, we leverage the full range of \planck data and make full use of the PR4 improvements. We also incorporate the \planck Commander low-$\ell$ temperature likelihood in our analysis, which remains based on PR3 as no significant improvements were expected in the PR4 intensity maps for this range of multipoles. The synergy between these likelihoods enables a robust estimation of both the reionisation parameter, $\tau$, from the large scales and the other cosmological parameters, in both temperature and polarisation. On the other hand, Appendix~\ref{app:lensing} demonstrates that incorporating information from the CMB-derived gravitational lensing power spectrum, while tightening the constraints on the amplitude of primordial scalar fluctuations ($A_{\rm s}$), has no effect on the reconstruction of the reionisation optical depth despite the known correlation between the two parameters. Consequently, the Planck CMB lensing likelihood is not included in our analysis.

Unless explicitly mentioned otherwise, we sampled the likelihoods for the free reionisation parameters along with the five parameters of the $\Lambda$CDM cosmological model (the physical baryon density, $\Omega_{\rm b}h^2$, the physical density of cold dark matter, $\Omega_{\rm cdm}h^2$, the amplitude of primordial scalar fluctuations, $A_{\rm s}$, the scalar index, $n_s$, and the Hubble constant, $H_0$), as well as the nuisance parameters associated with each likelihood. The latter are related to instrumental calibration and residual foreground modelling. We adopted the same priors on those nuisance parameters as described in~\citet{2024A&A...682A..37T}. Additionally, we assumed three degenerate neutrinos of fixed total mass 0.06\,eV.

\subsection{Tools}

Throughout our analysis, theoretical predictions for the CMB anisotropy power spectra were computed using the CLASS Boltzmann code \citep{2011JCAP...07..034B}. We developed a custom-modified version of CLASS\footnote{Publicly available at \href{https://github.com/s-ilic/class_reio}{github.com/s-ilic/class{\_}reio}}, which incorporates all the reionisation scenarios considered in this study. The only exception is the \texttt{pca-5} model, for which we use a modified version of CAMB \citep{Lewis:1999} as it allows for the computation of CMB power spectra even in cases where the ionisation fraction temporarily goes negative.

To perform cosmological inference, we used the ECLAIR suite of codes\footnote{Publicly available at \href{https://github.com/s-ilic/ECLAIR}{github.com/s-ilic/ECLAIR}}, designed as a versatile tool for sampling the posterior distribution of model parameters. ECLAIR \citep[first introduced in][]{Ilic:2020onu} integrates cutting-edge datasets, efficient affine-invariant ensemble MCMC algorithms \citep[via \texttt{emcee} and \texttt{zeus},][]{2013PASP..125..306F,karamanis2021zeus,karamanis2020ensemble}, and interfaces with the CAMB and CLASS Boltzmann solvers, or any custom modifications of them. The pipeline is highly parallelised, leveraging multiprocessing on a single machine or distributing computation with MPI, proving advantageous for the computationally demanding tasks associated with parameter space exploration. ECLAIR also includes a dedicated minimiser that combines ensemble sampling with simulated annealing, efficiently targeting the global maximum of any posterior. The convergence and robustness of the analyses are further supported by integrated scripts that facilitate validation of the estimated marginalised posterior distributions and credibility intervals, and take advantage of the capabilities of the \texttt{getdist} package \citep{2019arXiv191013970L} for outputting results.

We paid close attention to the convergence of the chains, especially for the more complex, weakly constrained models that exhibit very long correlation lengths (up to hundreds of steps). We ensured effective sample sizes of at least a few thousands (after burn-in), and Gelman-Rubin $R-1$ values below $0.01$ for all cosmological parameters.

\subsection{Handling of implicit priors}
\label{sec:priors}

Before diving into our results, it is essential to acknowledge that all reionisation models inherently carry a prior in the space of reionisation histories, $x_{\rm e}(z)$. Indeed, while we use uniform priors for all parameters of each reionisation model, this choice does not guarantee a uniform distribution of $x_{\rm e}$ values at any given $z$, and even less so across an interval of $z$ values. Consequently, this also imposes an implicit prior on the derived value of the optical depth, $\tau$, which can significantly influence its inferred posterior distribution and other reionisation-related quantities. As this $\tau$ prior is unique to each reionisation model, not accounting for it would bias any comparison between models since we would not be able to disentangle the effect of the prior versus what model the data actually prefer. We addressed this potential issue via two different approaches, described in the following sections. 

\subsubsection{Prior flattening}

One method consists in precisely characterising this implicit prior on $\tau$ in order to cancel its influence on the final posterior distribution. To do so, we consider the methodology proposed by \citet{Millea:2018}, which involves two key steps. First, for each reionisation model considered we sampled the prior distribution, $\mathcal{P}(\theta)$, of its parameters, $\theta$, and computed the corresponding marginalised distribution of derived $\tau$ values, $\mathcal{P}(\tau(\theta))$. This step allowed us to quantify any intrinsic bias on $\tau$ imposed by the chosen reionisation history priors.

\begin{figure}[!ht]
    \center
    \includegraphics[width=.9\columnwidth]{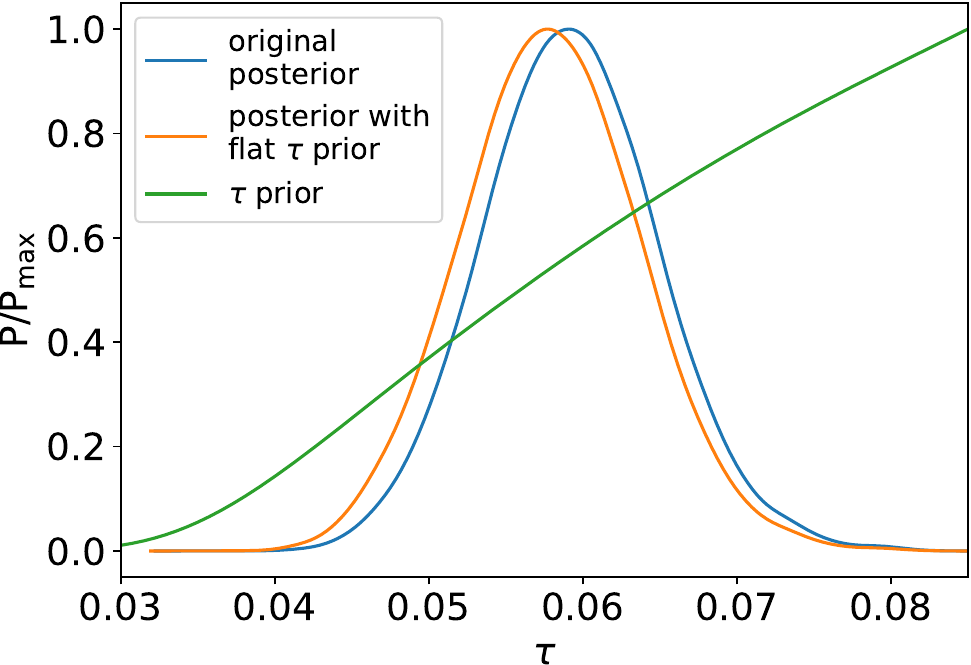}
    \caption{Illustration of the flattening process of the $\tau$ prior distribution, applied to the \texttt{3-steps} reionisation model. The original posterior on $\tau$ is shown in blue, while the posterior after flattening and the $\tau$ prior are shown, respectively, in orange and green.}
    \label{fig:flat-process}
\end{figure}

Next, we corrected for this prior-induced bias by importance sampling the points in our MCMC chains. More specifically, each MCMC sample was weighted by the inverse of the previously obtained distribution, $\mathcal{P}(\tau(\theta))$. This procedure ensures that our posterior distributions of $\tau$ are not artificially skewed by the initial choice of priors in the reionisation models. Written explicitly, we can create a posterior where we have assumed a flat $\tau$ prior by computing
\begin{align} \label{eq:flatten_tau}
    \mathcal{P}^{{\rm flat}\text{-}\tau}(\theta\,|\,d) = 
    \frac{\mathcal{P}(\theta\,|\,d)}{\mathcal{P}(\tau(\theta))} 
    \propto \mathcal{L}(\theta) \underbrace{\left[\frac{\mathcal{P}(\theta)}{\mathcal{P}(\tau(\theta))}\right]}_{\mathcal{P}^{{\rm flat}\text{-}\tau}(\theta)},
\end{align}
where the posterior given data $d$, $\mathcal{P}(\theta\,|\,d)$, is the original posterior that did not assume a flat $\tau$ prior; $\mathcal{P}^{{\rm flat}\text{-}\tau}(\theta\,|\,d)$ is the ``$\tau$-flattened'' posterior, and the likelihood, $\mathcal{L}(\theta)$, is the probability of the data given parameters, $\mathcal{P}(d\,|\,\theta)$. Figure~\ref{fig:flat-process} illustrates the ``flattening" process for a reionisation model whose $\tau$ posterior is noticeably affected, namely our \texttt{3-steps} model.

We note that there are an infinite number of ways to adjust the posterior distribution to achieve a flat prior on $\tau$. However, the aforementioned method ensures that the minimum amount of additional information is introduced into the analysis, or equivalently that the entropy of the prior-adjusted $\tau$ posterior is maximised (see Appendix A of \citealt{Millea:2018} for details). We further note that this method can be extended to any derived parameter of interest.

\begin{figure*}[!ht]
    \includegraphics[width=0.95\columnwidth]{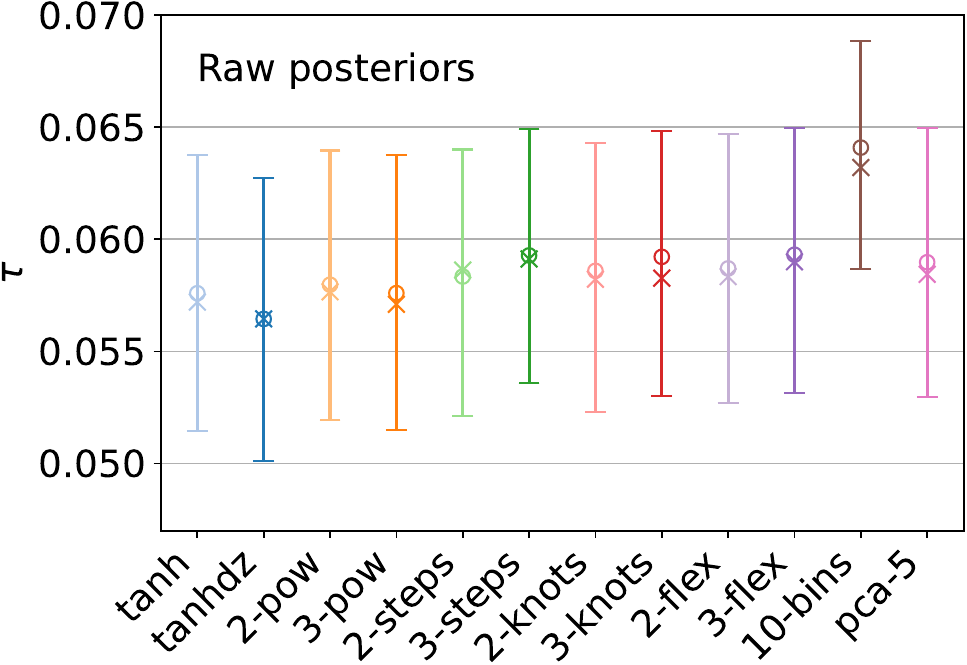} \hspace{0.05\columnwidth}
    \includegraphics[width=0.95\columnwidth]{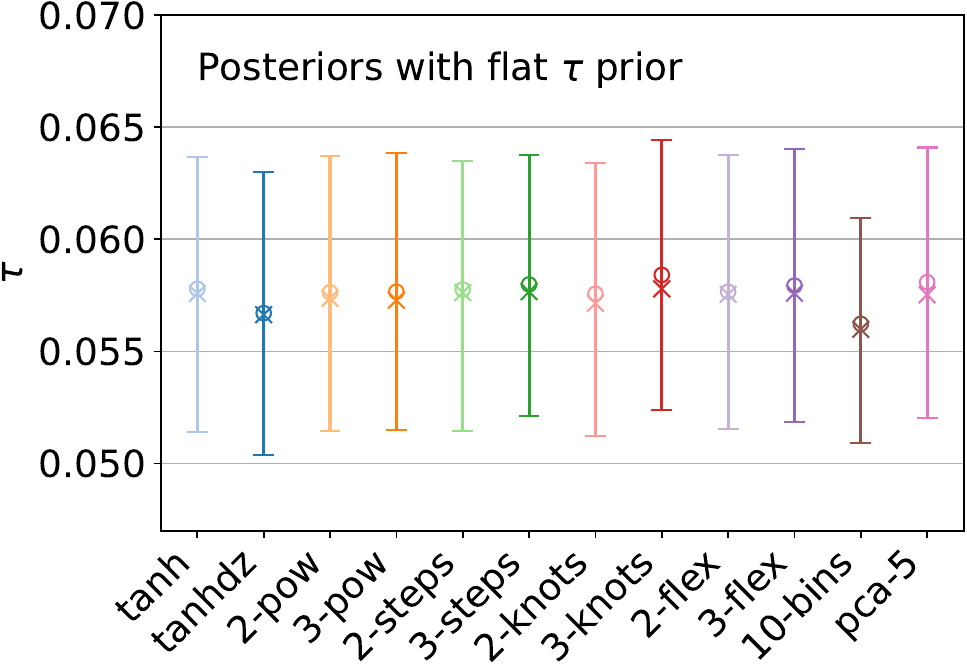}
    \caption{Constraints on the reionisation optical depth from \planck PR4 for the reionisation models considered. The plotted segments represent the 68\% credible intervals, where the crosses and circles mark, respectively, the maximum and mean of each posterior. The left panel shows the raw posterior results, while the right panel includes the prior-flattening correction described in Sect.~\ref{sec:priors}.}
    \label{fig:tau-summary}
\end{figure*}

\subsubsection{Profile likelihoods}

While the previously described method effectively removes the effects of prior assumptions on the final $\tau$ constraints, it does not address the further possibility of volume effects in the parameter space. Indeed, whole swaths of the oddly shaped parts of the parameter space can have similarly acceptable likelihood -- and $\tau$ values, which can then affect the posterior distribution of $\tau$ after marginalisation. This can be particularly pronounced for models with a large number of degrees of freedom. While such an effect is inherent to -- and expected from -- a Bayesian formalism, one may want to infer constraints without being sensitive to either the choice of priors or volume effects in the parameter space. To mitigate those effects, we consider an alternative approach based on profile likelihoods.

Using $\tau$ as an example, this method consists in successively fixing $\tau$ to a grid of predefined values and then minimising the chi-square for each distinct value of $\tau$, varying all other parameters in the analysis (both cosmological and nuisance-related). Selecting the resulting best-fit value and applying a standard chi-square threshold on the resulting likelihood profile then provides, respectively, a robust estimate of $\tau$ and its uncertainty, without the influence of any explicit or implicit prior \citep[see e.g.][for an application in the context of the cosmological analysis of the \planck CMB data]{planck2013-XVI}. We apply this method to the reionisation models considered in this work and compare the results to those obtained with the prior-flattening technique.

As a technical detail, we note that to perform the minimisations required by the profile likelihood, we had to reparameterise each of our models to include $\tau$ explicitly as an input parameter. For most models, this was simply achieved by introducing a parameter that can shift all redshift-related parameters at once, thus allowing for a direct control of $\tau$. For the \texttt{pca-5} and \texttt{10-bins} models, for the vector of model parameters, we switched from cartesian to spherical coordinates, and use $\tau$ to scale the norm of the aforementioned vector. Finally, we also had to increase the precision settings of the Boltzmann solvers to facilitate the minimisation procedure. Specifically, for CLASS we identified the most relevant parameters as \texttt{reionization\_sampling} (set to 0.01) and the \texttt{perturbations\_sampling\_stepsize} (set to 0.05), while for CAMB we set the \texttt{AccuracyBoost} parameter to 3.

\begin{figure*}[!ht]
    \center
    \includegraphics[width=.95\columnwidth,height=170pt,valign=t]{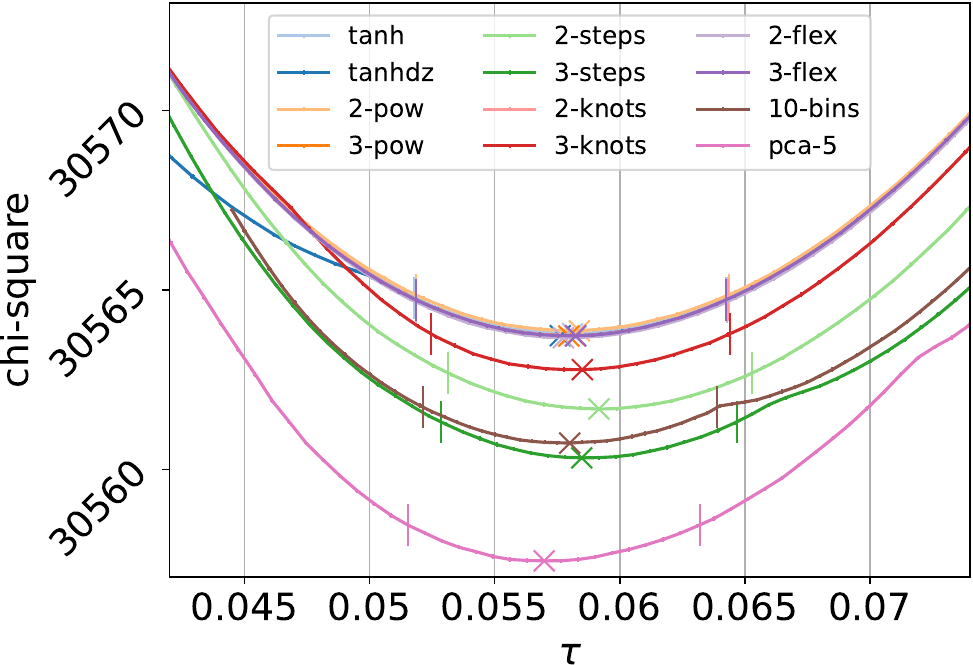}
    \hspace{0.05\columnwidth}
    \includegraphics[width=.95\columnwidth,valign=t]{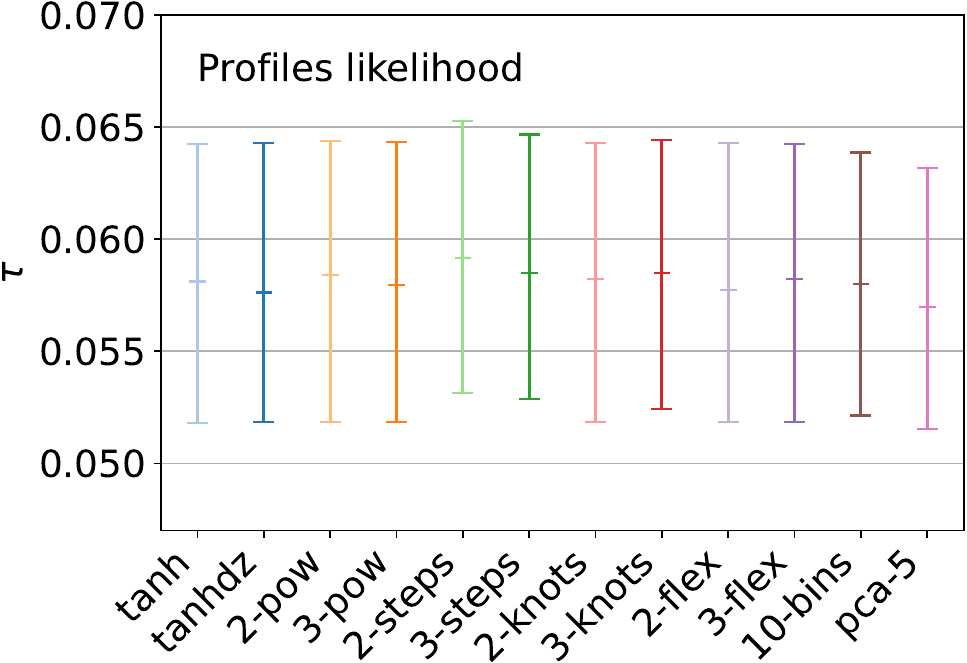}
    \caption{Left panel: Profile likelihoods of the reionisation optical depth, $\tau$, for the reionisation models considered, where crosses and small vertical bars, respectively, mark the maximum-likelihood values and the $\Delta \chi^2 < 1$ intervals. Right panel: Summary of the maximum likelihood values and $\Delta \chi^2 < 1$ intervals derived from the profiles, plotted as segments.}
    \label{fig:tau-profiles}
\end{figure*}

\section{Results on the reionisation optical depth}
\label{sec:results_tau}

In this section, we present our measurements of the reionisation optical depth, $\tau$, obtained as a derived parameter from our various reionisation models, as constrained by the \planck PR4 data. In our work we define $\tau$ as the integral defined in Eq.~(\ref{eq:tau_def}) with bounds set from $z=0$ to $z=30$.\footnote{We note that, by default, the upper redshift bound used by the CLASS code is set as the redshift of the global minimum of the ionisation history. We chose to override this definition in order to accommodate our potentially non-monotonic models (\texttt{3-steps} and \texttt{10-bins}), where such a minimum is not necessarily the start of reionisation.} We emphasize that all $\tau$ measurements reported below include an identical contribution (about 0.03) from $z<5$ resulting from the fixed timing and duration of the second helium reionisation.

\subsection{Posterior distributions}

For each of the 12 models considered in this work, we derived the posterior distribution of the reionisation optical depth from our corresponding MCMC chains. We summarise our results in Fig.~\ref{fig:tau-summary}, showing the obtained means and maxima of the posteriors as well as the 68\% credible intervals. The left panel of the figure displays the raw posteriors, while the right panel shows the posteriors after applying the prior-flattening correction described in Sect.~\ref{sec:priors}.

We first notice that the measurements of $\tau$ are all consistent between the different models. The raw posteriors of $\tau$ (left panel of Fig.~\ref{fig:tau-summary}) exhibit some relatively small dispersion with the exception of the \texttt{10-bins} model that has a higher value compared to the others by approximately 1$\sigma$. This aligns with expectations regarding the implicit prior effect, which is anticipated to be more pronounced in higher-dimensional parameter spaces, especially in cases such as ours where the data are not particularly constraining.

This is further confirmed by the results obtained from the posteriors after applying the prior-flattening correction (right panel of Fig.~\ref{fig:tau-summary}), which show better agreement between the different models.
This alignment indicates that the correction of the prior effect effectively mitigates the bias induced by model-specific priors, providing a more consistent estimate of $\tau$.

This consistency not only reinforces the credibility of our $\tau$ estimates but also reinforces the reliability of other cosmological parameters. Indeed, our findings indicate that the modelling of the reionisation history has no impact on the constraints for the other $\Lambda$CDM parameters after applying the $\tau$ prior-flattening procedure (see Appendix~\ref{app:lcdm}).

We also note that the width of the posteriors is quite stable across  models, despite significant differences in numbers of degrees of freedom (dof). The \planck PR4 data thus appear to be able to constrain $\tau$ with a similar precision, regardless of the complexity of the reionisation model. We may hypothesise about the origin of this observation: as briefly mentioned in Sect.~\ref{sec:Models}, the constraining power of the CMB on reionisation is two-fold: scalar spectra are damped by $e^{-2\tau}$ and the low-$\ell$ polarisation signal scales as $\tau^2$ (at first order). Together, these two effects represent the bulk of the constraint on $\tau$ from the CMB, and are likely to dominate the total variance budget. However, the large-scale polarisation signal also exhibits additional multipole-dependent features depending on the precise details of the reionisation history, $x_{\rm e}(z)$, which translates into an additional constraint on $\tau$ but is likely to be of secondary order. This could explain why the width of the posteriors for $\tau$ is so stable and is not significantly affected by the details and complexity of the chosen reionisation model.

The only notable exception in terms of consistency in the constraints comes from the \texttt{10-bins} model that, despite the prior-flattening correction, still exhibits a different, slightly smaller (by $\sim$20\%) $\tau$ uncertainty compared to the other models. This may seem paradoxical, as the \texttt{10-bins} model has the largest number of degrees of freedom. However, even though the prior-flattening procedure removes any volume effect affecting $\tau$, it does not remove potential volume effects affecting the likelihood itself. If present, the profile likelihood method (see next section) should be able to identify such remaining effects.

We summarise the quantitative constraints on $\tau$ in all the cases considered in Table~\ref{tab:tau_results} for each model. Ultimately, from the posterior distributions obtained with \planck CMB data, we measure the following averaged reionisation optical depth:
\begin{align}
    \tau = 0.0576 \pm0.0060 \pm0.0006,
    \label{eq:tau_bayes}
\end{align}
where the first uncertainty accounts for statistical errors (average of the 68\% $\tau$ uncertainties across models) while the second is given by the dispersion over the reionisation history models (standard deviation of the mean $\tau$ value across models).

\subsection{Profile likelihoods}

We also computed the profile likelihoods for $\tau$ for each reionisation model considered in this work. The results are shown in Fig.~\ref{fig:tau-profiles} and reported in Table~\ref{tab:tau_results}. The minimisation procedure in ECLAIR enables the reconstruction of the full shape of the profile likelihood, from which we extract the global minimum and define the confidence interval by applying a $\Delta\chi^2=1$ threshold. 
Most profiles are observed to be roughly parabolic in shape. The simpler models in particular form a remarkably consistent group with nearly identical profiles. The more complex models allow marginally better chi-square values (differing by at most six units)\footnote{Considering the differences in numbers of dof, and using metrics such as the Akaike information criterion (AIC) or Bayesian information criterion (BIC), such chi-square differences are not big enough to make any model significantly more or less preferable than the others.}, while some exhibit deviations from quadratic behaviour in their shape. Indeed, complex models (including \texttt{tanhdz}, \texttt{pca-5}, \texttt{10-bins} and \texttt{3-steps}) introduce additional degrees of freedom, subtle dependencies on the parameter, non-linear parameter effects and potential multimodality, all of which can lead to more intricate profile likelihoods.

The $\tau$ constraints from the profile likelihoods display the same consistency across different reionisation models as observed in the MCMC-derived posteriors. The profile likelihoods also exhibit a stable width across the various models, confirming the robustness of our results. Interestingly, the profile likelihoods make the results on $\tau$ even more coherent, including for the \texttt{10-bins} model that was preferring slightly lower values of $\tau$ and tighter errors as noted in the previous section, thus confirming our hypothesis regarding the remaining volume effects.

Following the same prescription as for Eq.~(\ref{eq:tau_bayes}), we find as constraints on $\tau$ from profile likelihoods
\begin{align}
    \tau = 0.0581 \pm{0.0061} \pm 0.0005.
    \label{eq:tau_frequentist}
\end{align}

As was previously discussed, profile likelihoods are not affected by volume effects, thus any visible discrepancy between them and MCMC results are likely a consequence of these influences. The most notable impact is observed in the \texttt{10-bins} model (see Table~\ref{tab:tau_results}). The average value of $\tau$ from profile likelihoods is slightly higher (albeit by only $\sim0.1\sigma$), while the uncertainties become noticeably more consistent.

\begin{figure*}[!ht]
    \center
    \includegraphics[width=.95\textwidth]{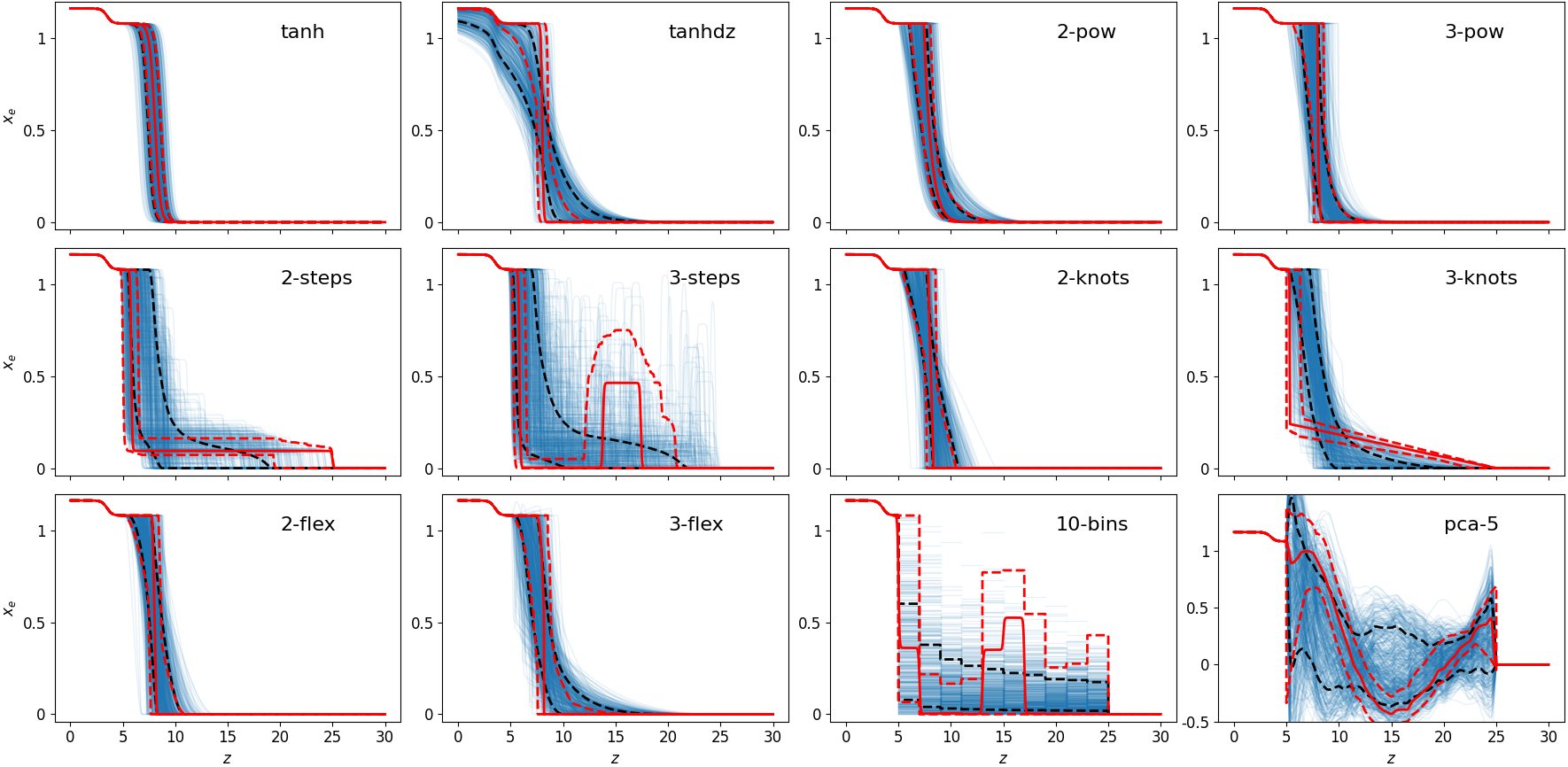}
    \caption{Ionisation history as a function of the redshift, $x_{\rm e}(z)$, for the reionisation models considered. A few hundred samples from the MCMC distributions are shown in blue, while dashed black lines represent the 68th percentile of the distribution of $x_{\rm e}(z)$ for each redshift $z$. The solid and dashed red lines, respectively, correspond to the best-fit solutions and the frequentist confidence intervals determined by $\Delta\chi^2<1$.}
    \label{fig:xe-summary}
\end{figure*}

\section{Results on the history of reionisation}
\label{sec:results_xe}

In this section, we examine in more detail the constraints on the evolution of the ionised fraction of matter itself as a function of redshift as well as associated derived quantities.

\subsection{Redshift evolution of the ionisation fraction}

We explore the reconstructed reionisation histories for all the models considered in Sect.~\ref{sec:Models}. In each case, we plot $x_{\rm e}(z)$, providing a visual representation of the constraints on the evolution of the ionisation fraction over cosmic time. Each reionisation model, from the classic hyperbolic tangent to the flexknot approach, yields distinct $x_{\rm e}(z)$ curves that reflect their underlying assumptions and parametrisations. These reconstructions allow us to assess the flexibility and accuracy of each model in fitting the \planck PR4 data, thereby offering insights into the nature and duration of the reionisation process.

We summarise the reconstruction results in Fig.~\ref{fig:xe-summary}, which displays in blue the MCMC distributions of histories of the ionisation fraction as a function of redshift, $x_{\rm e}(z)$, for all models considered in this work. The blue curves envelop the 68\% credible intervals at each $z$. Given that the sampling of the reionisation models is performed using flat priors on the parameters, this consequently induces non-flat priors on $x_{\rm e}(z)$. As a result, their posterior distributions may be significantly influenced by prior effects, which can vary substantially between models. Similarly to the previous discussion on the reionisation optical depth (Sect.~\ref{sec:results_tau}), we also derived the best-fit history for each model, shown as a solid red line. The two dashed red lines mark the boundaries of the region containing models that fall within $\Delta\chi^2=1$ of the best-fit value, illustrating the confidence interval associated with the best-fit. We note that, as a consequence of the prior effects, combined with complex likelihood shapes associated with the most sophisticated models, the best-fit $x_{\rm e}(z)$ can fall outside of the 68\% credible intervals.

We observe that the simpler models, with less than 3 degrees of freedom (dof), all exhibit a similar behaviour. These models include \texttt{tanh}, \texttt{2-pow}, \texttt{2-knots}, \texttt{2-flex}, \texttt{3-pow}, and to a lesser extent \texttt{tanhdz}. They are largely consistent with an instantaneous reionisation scenario, resembling the classic hyperbolic tangent model. This shows that when the model flexibility is limited, both the MCMC exploration and best-fit solutions tend to favour a rapid reionisation history.

In contrast, the intermediate complexity models, with up to 5 dof, show a more nuanced picture. These models include \texttt{2-steps}, \texttt{3-knots}, and \texttt{3-flex}. They consistently exhibit a tail towards higher redshifts, indicating the possibility of a moderate contribution to reionisation at early times (z > 10). This tail is coherent across different models, suggesting that the data can support the presence of a limited ionisation fraction at higher redshifts.

Finally, the highest complexity models (such as \texttt{3-steps}, \texttt{pca-5}, and \texttt{10-bins}), offer an even more detailed insight. These models are consistent with the intermediate-complexity models, but we note that their best-fit consistently suggests a non-zero ionisation fraction around redshift $z\sim15$. This peak, while moderate in significance, suggests that reionisation history may have been more complex than expected. We note that, by construction, the \texttt{pca-5} model allows the ionisation fraction to reach values beyond the physical range $[0,1+f_{\rm He}]$. This occurs because the completeness of the orthogonal basis is broken when limiting the number of modes, even more so in our case where we only use five (see Sect.~\ref{sec:Models:pca}).

As outlined in Sect.~\ref{sec:results_tau}, the constraints on the ionisation fraction derived from the CMB power spectra are primarily driven by the optical depth. In practice, the reconstructed ionisation histories all conform to the same $\tau$, within the flexibility permitted by each model.  More complex models can extract additional information from the detailed shape of the $C_\ell^{EE}$ spectrum, providing a further constraining power -- though its statistical significance remains limited. We present in Appendix~\ref{app:cl_bestfit} the E-mode CMB polarisation angular power spectra ($C_\ell^{EE}$) corresponding to the best-fit of each reionisation history model. Notably, models that exhibit a non-zero ionisation fraction at $z\sim15$ show an increase in power at scales around $\ell = 6-25$ (compared to the more instantaneous models), which corresponds to the region just after the reionisation bump. While this additional power improves the fit to the data, its statistical significance is not high enough to determine whether it originates from a feature of the reionisation history. Indeed, in this multipole range, the signal-to-noise ratio of \planck is below one, and the data may be affected by systematic residuals (arising from foreground or instrumental contamination).

\subsection{Constraints on the end of reionisation}

In this section, we explore to what extent the CMB could place constraints on the end of reionisation. To investigate this, we extend the \texttt{10-bins} model by introducing two additional bins at very low redshifts: one spanning $z=0$ to $z=2.5$, and the other from $z=2.5$ to $z=5$. Unlike previous models, we do not impose a second reionisation of helium but instead allow the first two bins to vary freely between 0 and $1+2f_{\rm He}$.

\begin{figure}[!ht]
    \center
    \includegraphics[width=.95\columnwidth]{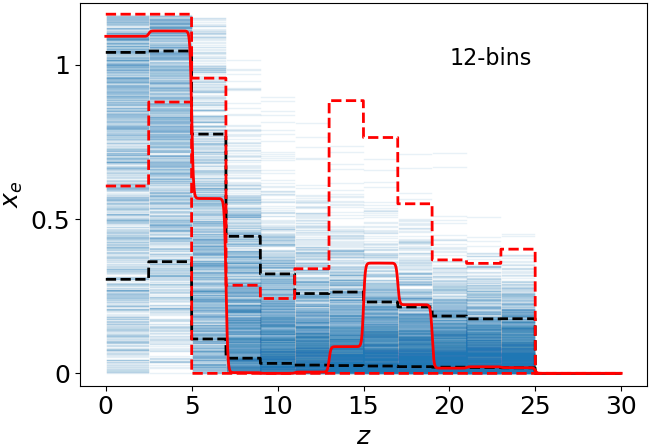} 
    \caption{Ionisation history as a function of the redshift, $x_{\rm e}(z)$, for the \texttt{12-bins} model. The dashed black lines represent the 68th percentile of the distribution of $x_{\rm e}(z)$ for each redshift $z$. The red lines correspond to the best-fit solution with the corresponding 1-$\sigma$ confidence interval given by the dashed red lines.}
    \label{fig:xe-12-bins}
\end{figure}

The resulting ionisation fraction as a function of redshift is shown in Fig.~\ref{fig:xe-12-bins}. The recovered reionisation history closely resembles that of the \texttt{10-bins} model, with a global transition occurring between $z=7.5$ and $z=5$. As in the \texttt{10-bins} case, the best-fit solution for the \texttt{12-bins} model also shows a non-zero ionisation fraction around $z=15$. Notably, in contrast to the higher redshift bins, we find a preference for a non-zero ionisation fraction in the first two bins, with both a peak of the $x_{\rm e}$ posterior distribution and a best-fit $x_{\rm e}$ value close to $1+2f_{\rm He}$.

The optical depth values recovered for this \texttt{12-bins} model are $\tau = 0.0548 \pm 0.0069$ (posterior with flat-$\tau$ prior) and $\tau = 0.0575_{-0.0072}^{+0.0076}$ (profile likelihood). In this case, the mean posterior distribution of $\tau$ is slightly lower than in the other models, as this model allows for incomplete reionisation at $z=0$, a scenario that was excluded in previous models except for the \texttt{tanhdz} one (which also exhibited such slight shift towards lower $\tau$). However, the best-fit value of $\tau$ remains fully consistent with the one obtained from the \texttt{10-bins} model.

\begin{figure*}[!ht]
    \center
    \includegraphics[width=.9\textwidth]{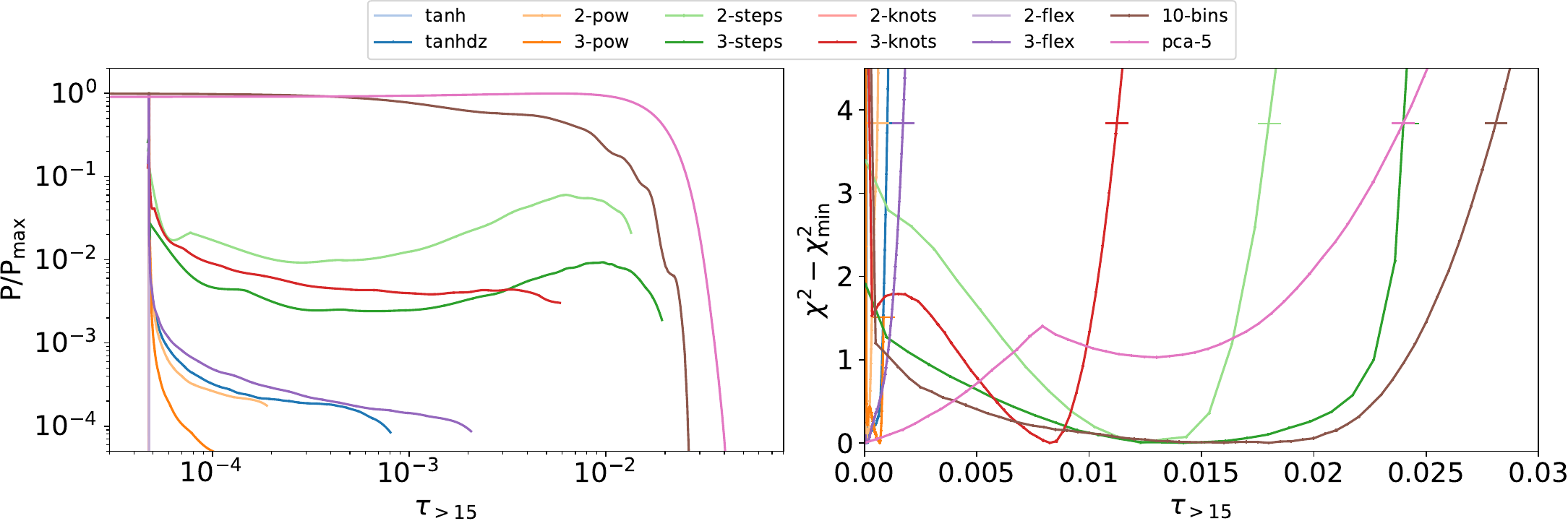}
    \caption{Constraints on $\tau_{>15}$ for the reionisation models considered. Left panel: full posterior distributions, with a flattening procedure applied to the $\tau_{>15}$ prior. Right panel: profile likelihoods for $\tau_{>15}$.}
    \label{fig:tau_15-30_posteriors}
\end{figure*}

\subsection{Constraints on early sources of reionisation}

In addition to constraining the ionisation fraction, $x_{\rm e}(z)$, we also derived estimates for redshift-integrated quantities such as the cumulative reionisation optical depth from $z$ to $z_{\rm max}$, $\tau_{[z,z_{\rm max}]}$ (see Appendix~\ref{app:tauz}). As noted by~\citet{Heinrich:2017}, this quantity is more directly connected to the polarisation observables than the instantaneous ionisation fraction, $x_{\rm e}(z)$. In our case, the shape of $\tau_{[z,z_{\rm max}]}$ below $z=5$ is fixed by construction. Similar to $x_{\rm e}(z)$, our results show that the contribution to optical depth at $z>10$ are highly model-dependent.

Specifically, we consider the optical depth integrated over the redshift range $z>15$. Indeed, this quantity is of particular interest for probing early reionisation signatures, as it highlights the contribution to $\tau$ from early ionising sources, prior to redshift 15. Figure~\ref{fig:tau_15-30_posteriors} displays the \planck PR4 constraints on $\tau_{>15}$ for the various models discussed in this work. The left panel presents the posterior distributions, while the right panel depicts the profile likelihoods estimated for $\tau_{>15}$. As with $\tau$, the posterior distribution for $\tau_{>15}$ is similarly influenced by prior choices. We switch to a flat prior on $\tau_{>15}$ after applying the same type of correction as for $\tau$ (see Sect.~\ref{sec:priors}).

The results for the simplest models (with less than 3 degrees of freedom) yield almost identical and tightly constrained values of $\tau_{>15}$, centred around $\sim 4.75 \times 10^{-5}$, consistent with the minimal residual ionisation remaining from the recombination era. This outcome is within expectations, as these models are designed to capture only the primary transition from a neutral to an ionised intergalactic medium, lacking the flexibility to model an additional, independent contribution to ionisation at earlier times. Indeed, for such models, higher values of $\tau_{>15}$ can only be achieved for sets of parameters that lead to a correspondingly larger total optical depths, which are disfavoured by the data.

The models that allow for a more complex transition with an early contribution at high redshift are also compatible with zero early contribution to reionisation. Only the posteriors for the most complex models, namely the \texttt{10-bins} and \texttt{pca-5}, exhibit a preference for a non-zero value even after prior flattening. However, the profile likelihoods on $\tau_{>15}$ for these complex models are less decisive and show a large region of equivalent statistical significance between $\tau_{>15} = 0$ and $0.02$. 

Based on the profile likelihoods, we derive 95\% confidence level upper limits for $\tau_{>15}$ by applying a purely data-based threshold (independent of priors or parameterisations) of ${\Delta\chi^2<3.841}$, which corresponds to the critical value of the chi-square distribution for 95\%, resulting in the following upper-bounds:
\begin{align}
    \tau_{>15} < 0.028  &\quad \text{(\texttt{10-bins})} \, , \\
    \tau_{>15} < 0.024  &\quad \text{(\texttt{3-steps})} \, , \\
    \tau_{>15} < 0.024  &\quad \text{(\texttt{pca-5})} \, , \\
    \tau_{>15} < 0.018  &\quad \text{(\texttt{2-steps})} \, , \\
    \tau_{>15} < 0.011  &\quad \text{(\texttt{3-knots})} \, ,
\end{align}
while the upper-bound is lower than $0.002$ for the other models, including \texttt{tanh}, \texttt{tanhdz}, \texttt{2-pow}, \texttt{3-pow}, \texttt{2-knots},  \texttt{2-flex}, \texttt{3-flex}. This highlights the challenges in measuring potential contributions from early sources of reionisation with CMB data. Our results show that the constraints on $\tau_{>15}$ are strongly model-dependent. Our upper-bounds are compatible with the results derived from previous Planck data sets with specific models (cf. $\tau_{>15}<0.020$ with \texttt{pca-5}, \citealp{Heinrich:2021}; $\tau_{>15}<0.018$ with FlexKnot, \citealp{planck2016-l06b}).

\section{Conclusions}
\label{sec:Conclusions}

In this work, we have reconstructed constraints on the history of cosmic reionisation using the measurements of CMB anisotropies, systematically exploring a wide range of reionisation models, from simple parametric forms to more flexible, non-parametric approaches. By employing both Bayesian and frequentist methods, we have carefully examined the impact of model-dependent priors and volume effects on the inferred constraints.

Our work highlights the importance of accounting for implicit priors when interpreting CMB-based constraints on reionisation. Different modelling approaches can introduce substantial biases, particularly for models with higher degrees of freedom, where prior effects become more pronounced. By correcting for these biases using prior-flattening techniques and complementing MCMC posteriors with profile likelihoods, we have ensured a more robust and unbiased interpretation of the results.

Our analysis confirms that \planck PR4 data provide robust constraints on the integrated optical depth, $\tau$, reaching an accuracy of approximatively 10\%. We find consistent $\tau$ estimates across the various reionisation models considered. Assuming a flat prior on $\tau$, we obtain $\tau = 0.0576 \pm 0.0060$, to which we add an additional uncertainty of around $0.0006$ to account for reionisation history modelling. In addition to prior effects, we clearly identify the presence of volume effects in the distribution of reionisation parameters. While their impact remains limited with \textit{Planck} data, such effects will require careful consideration in future CMB surveys with higher precision, where they could play a more significant role in shaping parameter constraints.

Despite consistent estimates of the optical depth, the reconstruction of the ionisation history, $x_{\rm e}(z)$, remains strongly model-dependent, as the CMB primarily constrains the integrated optical depth rather than the detailed redshift evolution of reionisation. Simpler models, which impose tighter assumptions, typically favour a rapid and relatively late ($z<10$) reionisation process, closely matching the standard hyperbolic tangent scenario. In contrast, more flexible models accommodate a wider variety of histories, including some early ionisation at $z\sim15$. However, this feature does not appear in the Bayesian posterior distributions and its significance remains below the 1-$\sigma$ level. As such, it cannot be determined from current data whether this early ionisation signal is physically motivated or simply a result of model flexibility.

Building upon the model-dependent nature of the reconstructed $x_{\rm e}(z)$, we further explored the potential for early ionisation by examining the parameter $\tau_{>15}$, which isolates the contribution to the optical depth from redshifts $z>15$. Among all models considered, only the more flexible \texttt{10-bins} model shows a slight preference for non-zero values in the posterior distribution. However, the corresponding profile likelihood remains broad, reflecting significant degeneracies and reaffirming that current CMB data alone lack the constraining power to robustly determine the presence of early ionising sources. From the profile likelihoods, we derived upper-bounds on $\tau_{>15}$, which are highly model-dependent, ranging from negligible up to ${\tau_{>15}<0.028}$ depending on the model considered.

Overall, our results show that the low value of $\tau$ measured from CMB data could be naturally reconciled with a high-redshift onset of reionisation suggested by recent JWST observations, by adopting more complex reionisation histories that allow for a slow and extended initial phase of ionisation.

While \planck PR4 data represent the most precise CMB measurements to date, further progress in understanding the EoR will require complementary observations. Next-generation CMB experiments, such as LiteBIRD~\citep{Litebird:2023} or CLASS~\citep{Watts:2018}, will significantly improve the precision of optical depth measurements through their enhanced sensitivity to large-scale polarisation anisotropies. This will help to break parameter degeneracies -- such as with the sum of neutrino masses -- and provide a more accurate anchor for the reionisation timeline. In particular, more accurate measurements of the large-scale CMB anisotropies in the $\ell = 10-25$ range will help constraining possible ionisation onset at high redshift \citep{Watts:2020}. However, our study shows that while current CMB data can robustly constrain the overall integrated optical depth, achieving a complete understanding of the redshift-dependent evolution of reionisation remains very challenging. This difficulty arises because the CMB spectrum is only weakly sensitive to the detailed shape of the reionisation history, leading to strong parameter degeneracies when fitting the latter to CMB data alone. Future observations of the 21cm power spectrum, such as HERA or SKA, will be crucial in directly mapping the spatial distribution of neutral hydrogen at high redshifts~\citep{2010ARA&A..48..127M,2012RPPh...75h6901P, MellemaKoopmans_2013,deBoer:2017}. The combination of upcoming CMB with 21 cm and UV luminosity observations will enable tighter constraints on the timing, duration, and morphology of reionisation, helping to identify the dominant ionising sources that shaped this pivotal phase of cosmic history \citep{Qin:2020,Paoletti:2021}.

\begin{table*}[!ht]
    \renewcommand{\arraystretch}{1.4}
    \caption{Constraints on the optical depth, $\tau$, for the reionisation models considered}
    \centering
    \begin{tabular}{llcccc}
    Model &  & dof & \multicolumn{3}{c}{$\tau$ 68\% constraints}\\
     &  &  & raw posterior & posterior with flat-$\tau$ prior & profile likelihood\\
    \hline
Hyperbolic tangent
    & \texttt{tanh}    &  1 & $0.0576\pm 0.0061$           & $0.0578^{+0.0059}_{-0.0064}$ & $0.0581_{-0.0063}^{+0.0062}$ \\
    & \texttt{tanhdz}  &  2 & $0.0564\pm 0.0063$           & $0.0567\pm 0.0063$           & $0.0576_{-0.0058}^{+0.0067}$ \\ \hline
Power-law
    & \texttt{2-pow}   &  2 & $0.0580^{+0.0060}_{-0.0061}$ & $0.0576^{+0.0061}_{-0.0062}$ & $0.0584_{-0.0066}^{+0.0060}$ \\
    & \texttt{3-pow}   &  3 & $0.0576^{+0.0062}_{-0.0061}$ & $0.0577\pm 0.0062$           & $0.0580_{-0.0061}^{+0.0064}$ \\ \hline
Step-based
    & \texttt{2-steps} &  3 & $0.0583^{+0.0057}_{-0.0062}$ & $0.0578^{+0.0058}_{-0.0063}$ & $0.0592_{-0.0060}^{+0.0061}$ \\
    & \texttt{3-steps} &  5 & $0.0593^{+0.0056}_{-0.0057}$ & $0.0580^{+0.0058}_{-0.0059}$ & $0.0585_{-0.0056}^{+0.0062}$ \\ \hline
Interpolation-based
    & \texttt{2-knots} &  2 & $0.0586^{+0.0057}_{-0.0063}$ & $0.0576^{+0.0058}_{-0.0063}$ & $0.0582_{-0.0064}^{+0.0061}$ \\
    & \texttt{3-knots} &  4 & $0.0592^{+0.0056}_{-0.0062}$ & $0.0584\pm 0.0060$           & $0.0585_{-0.0061}^{+0.0059}$ \\
    & \texttt{2-flex}  &  2 & $0.0587\pm 0.0060$           & $0.0577\pm 0.0061$           & $0.0578_{-0.0059}^{+0.0065}$ \\
    & \texttt{3-flex}  &  4 & $0.0593^{+0.0057}_{-0.0062}$ & $0.0579\pm 0.0061$           & $0.0582_{-0.0064}^{+0.0060}$ \\ \hline
Binned
    & \texttt{10-bins} & 10 & $0.0641^{+0.0047}_{-0.0054}$ & $0.0562^{+0.0047}_{-0.0053}$ & $0.0580\pm0.0059$ \\
    \hline
PCA
    & \texttt{pca-5}   &  5 & $0.0590\pm 0.0060$           & $0.0581\pm 0.0060$           & $0.0570_{-0.0055}^{+0.0062}$ \\
    \hline
Average
    &    &   &     & $0.0576 \pm 0.0060 \pm 0.0006$  & $0.0581 \pm 0.0061 \pm 0.0005$ \\
    \end{tabular}
    \label{tab:tau_results}
\end{table*}

\section*{Data availability}

All data products generated in the course of this analysis are publicly available on Zenodo at \href{https://doi.org/10.5281/zenodo.15706271}{10.5281/zenodo.15706271}.

\begin{acknowledgements}
    The authors acknowledge the support of the French Agence Nationale de la Recherche (ANR), under grant ANR-22-CE31-0010 (project BATMAN). 
    This work was supported by the French national space agency (CNES). 
    This work was performed using computational resources from the ``M\'esocentre'' computing centre of Universit\'e Paris-Saclay, CentraleSup\'elec and \'Ecole Normale Sup\'erieure Paris-Saclay supported by CNRS and R\'egion Île-de-France.\footnote{\href{https://mesocentre.universite-paris-saclay.fr}{mesocentre.universite-paris-saclay.fr}}
\end{acknowledgements}

\bibliographystyle{aa}
\bibliography{References,Planck_bib}

\begin{appendix}

\section{Effects of adding lensing likelihood}
\label{app:lensing}

\begin{figure}[!ht]
    \center
    \includegraphics[width=.95\columnwidth]{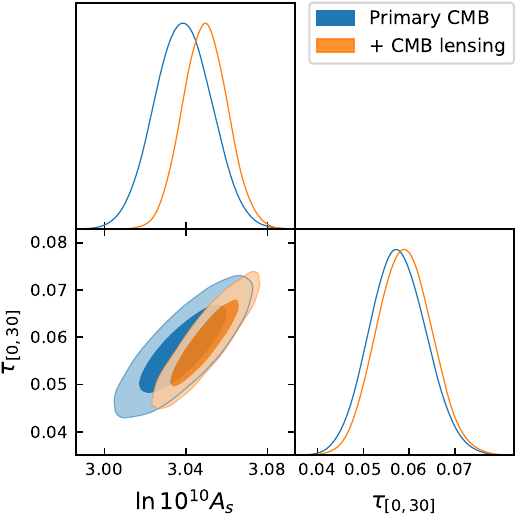}
    \caption{Effects of adding the \planck CMB lensing likelihood to the \texttt{tanh} model constraints on $\tau$ and $A_{\rm s}$. The contours show the 68\% and 95\% credible regions for the parameters.}
    \label{fig:tau-lensing}
\end{figure}

In Fig.~\ref{fig:tau-lensing}, we show the impact of the inclusion of the latest lensing likelihoods from \planck PR4 \citep{Carron:2022} and ACT DR6 \citep{2024ApJ...962..113M,2024ApJ...962..112Q} on constraints for the \texttt{tanh} reionisation model. As expected, adding lensing tightens the constraints on the amplitude of the primordial power spectrum, $A_{\rm s}$, reflecting its sensitivity to the matter distribution at intermediate redshifts ($z\sim$~1~--~5). However, it has negligible effect on the marginalised constraints on the reionisation optical depth, $\tau$, with nearly identical 68\% confidence intervals compared to analyses without lensing. While the lensing power spectrum, $C_\ell^{\phi\phi}$, is almost independent of $\tau$, the lensing likelihood depends on the lensed CMB spectra through the bias terms ($N^{(0)}$, $N^{(1)}$), mean-field, and estimator normalisation. In practice, with the current noise level, the $\tau$--$A_{\rm s}$ degeneracy remains mostly aligned with that from primary CMB anisotropies, so lensing does not significantly help in breaking it.

\section{Angular power spectra}
\label{app:cl_bestfit}

\begin{figure}[!ht]
    \center
    \includegraphics[width=.95\columnwidth]{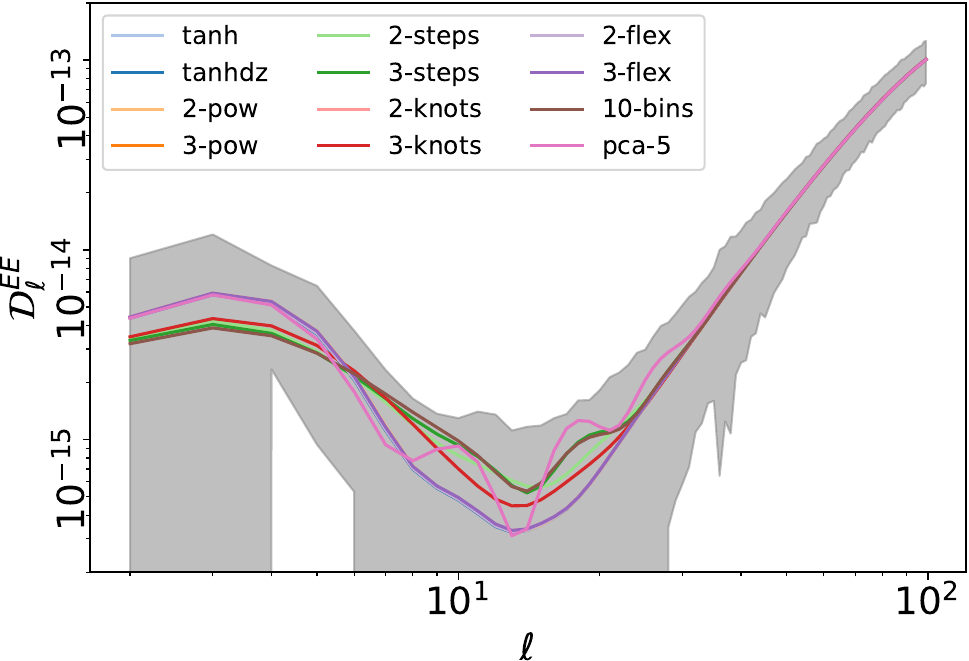}
    \caption{Angular power spectra of the CMB E-mode polarisation for the best-fit set of parameters for each reionisation model considered. The grey band indicates the \planck PR4 data full error bars (including sample variance, instrumental noise and systematics), centred on the \texttt{tanh} best-fit.}
    \label{fig:clEE_bestfit}
\end{figure}

In this appendix, we discuss the impact of the reionisation history on the CMB E-mode polarisation angular power spectra. Figure~\ref{fig:clEE_bestfit} shows $D_\ell = \ell(\ell+1)C_\ell^{EE}/2\pi$ for the best-fit corresponding to each reionisation model described in the main text. We observe that models with a greater number of degrees of freedom, which allow for a more extended reionisation history, consistently exhibit increased power at multipole scales between $\ell=5$ and $\ell=25$. On the contrary, in these models, the reionisation bump appears lower in amplitude, maintaining a consistent total reionisation optical depth across all models.

We note on the other hand that the $TT$ and $TE$ power spectra do not change significantly across the various best-fit reionisation history, as they are primarily sensitive to the amplitude of the primordial power spectrum and the reionisation optical depth.

\section{Cosmological parameters}
\label{app:lcdm}

\begin{figure}[!ht]
    \center
    \includegraphics[width=.95\columnwidth]{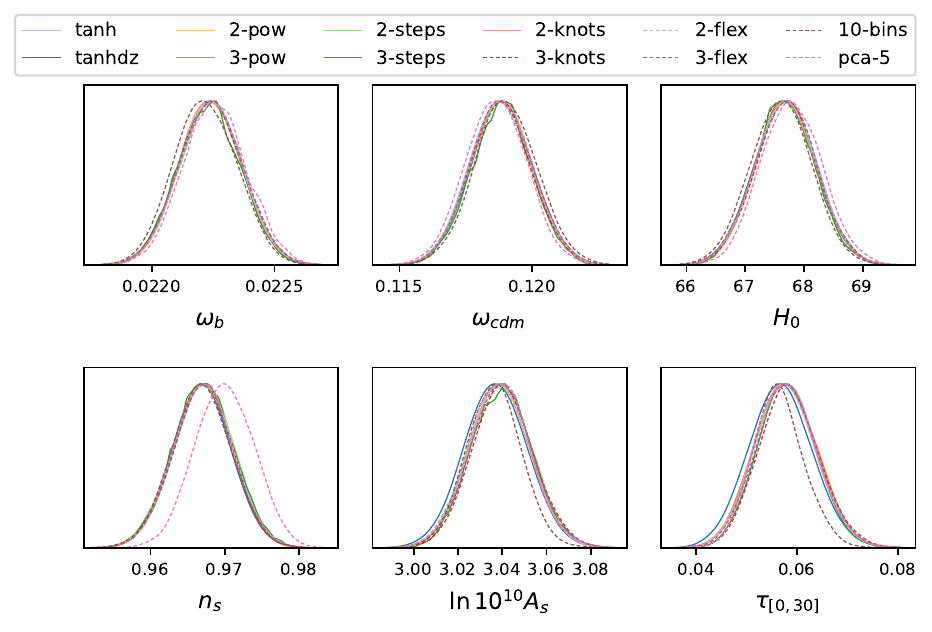}
    \caption{Posterior distributions for the $\Lambda$CDM cosmological parameters, including the reionisation optical depth, $\tau$, as a derived parameter, obtained from \planck PR4 data for the reionisation models considered.}
    \label{fig:LCDM}
\end{figure}

Figure~\ref{fig:LCDM} presents the posterior distributions for the $\Lambda$CDM parameters obtained when fitting the different reionisation models explored in this study, after the flattening procedure has been applied to the $\tau$ prior. These parameters include the physical baryon density, $\Omega_{\rm b}h^2$, the physical density of cold dark matter, $\Omega_{\rm cdm} h^2$, the amplitude of primordial scalar fluctuation, $A_{\rm s}$, the scalar index, $n_{\rm s}$, and the Hubble constant, $H_0$. We also include the reionisation optical depth, $\tau$, derived from the reconstructed reionisation history. Our results indicate that the inferred constraints on these parameters are robust and largely unaffected by the specific reionisation model assumed. The only exception is $n_{\rm s}$, which shows a slight shift towards higher values in the case of the \texttt{pca-5} model. This may be due to minor differences between CLASS and CAMB, as the latter is used to compute the spectra for the \texttt{pca-5} model. However, we were able to reproduce identical parameter distributions -- including $n_s$ -- for the \texttt{tanh} model using both codes, suggesting that the shift is specific to the \texttt{pca-5} implementation rather than the choice of Boltzmann solver.

\onecolumn

\section{Cumulative optical depth}
\label{app:tauz}

\begin{figure*}[!h]
    \includegraphics[width=\textwidth]{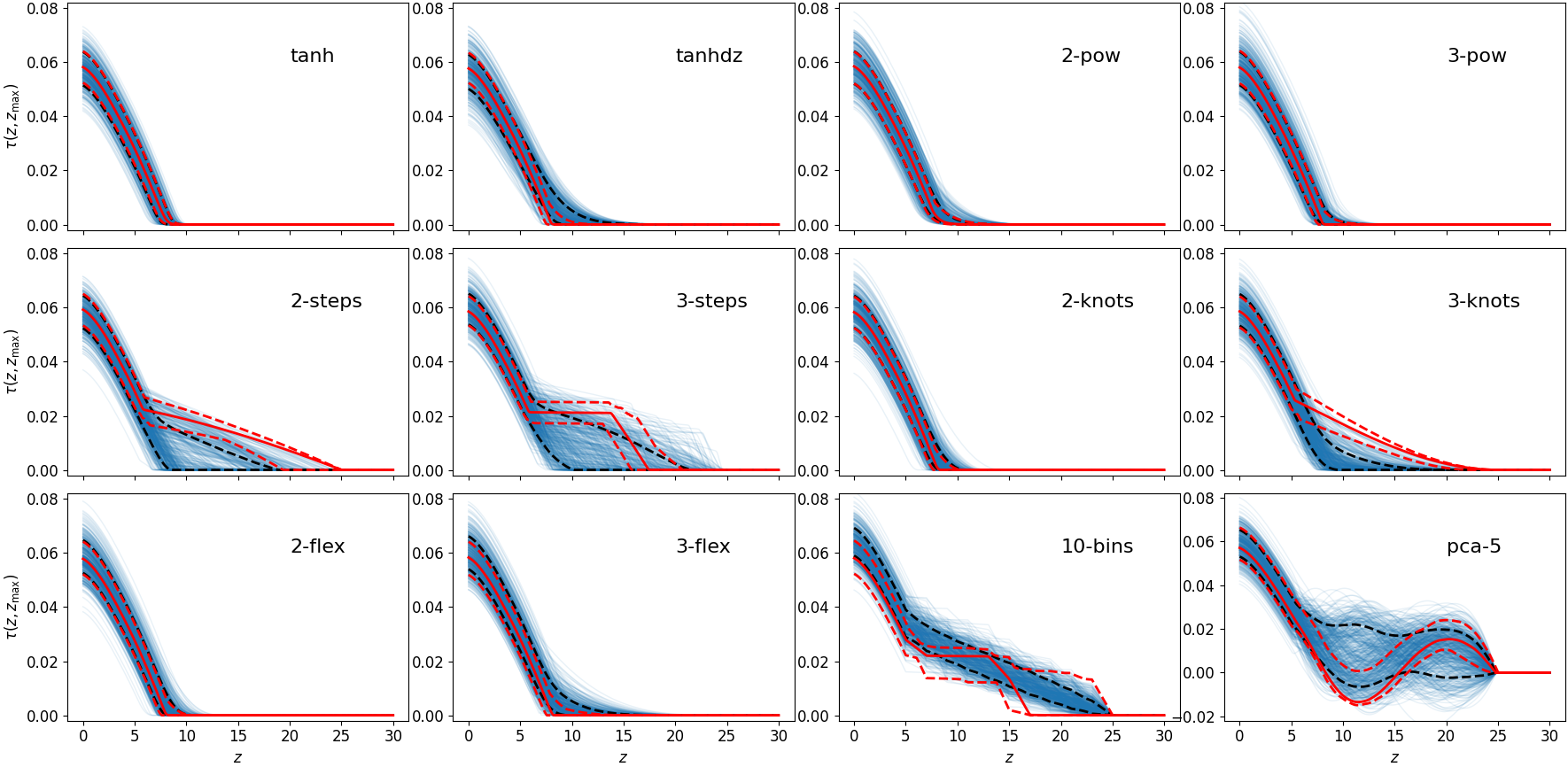}
    \caption{Cumulative optical depth (68\% and 95\% C.L. contours) for the reionisation models considered. The dashed black lines represent the 68th percentile of the distribution of $\tau(z)$ for each redshift $z$. The red lines correspond to the best-fit solution (solid) with the corresponding 1-$\sigma$ confidence interval (dashed).}
    \label{fig:tauz-summary}
\end{figure*}

The cumulative optical depth quantifies the integral of the electron density weighted by the Thomson cross section along the line of sight, and is calculated between redshifts $z$ and $z_{\rm max}$ as:
\begin{equation}
    \tau(z,z_{\rm max}) = \int_{z}^{z_{\rm max}} \sigma_T n_{\rm e}(z) \frac{dr}{dz} dz \, .
    \label{eq:tauz_def}
\end{equation}
By definition, the total optical depth is given by $\tau = \tau(0,z_{\rm max})$. Figure~\ref{fig:tauz-summary} displays the reconstructed $\tau(z,z_{\rm max})$ for each reionisation model considered in this work. Because we fixed the ionisation fraction, $x_{\rm e}(z)$, up to $z=5$, the shape of $\tau(z,z_{\rm max})$ below $z = 5$ is fixed by construction. Our results indicate that the contribution to optical depth at $z > 10$ are highly model-dependent. Among models with enough flexibility to allow some contribution to $\tau$ at high redshifts, the \texttt{2-steps}, \texttt{3-steps},  \texttt{3-knots}, \texttt{10-bins} and \texttt{pca-5} show non-zero contribution at $z>15$.

\end{appendix}

\end{document}